# Risk Assessment with Generic Energy Storage under Exogenous and Endogenous Uncertainty

Ning Qi, *Student Member, IEEE,* Lin Cheng, *Senior Member, IEEE,* Yuxiang Wan, Yingrui Zhuang and Zeyu Liu, *Student Member, IEEE*

*Abstract*—Current risk assessment ignores the stochastic nature of energy storage availability itself and thus lead to potential risk during operation. This paper proposes the redefinition of generic energy storage (GES) that is allowed to offer probabilistic reserve. A data-driven unified model with exogenous and endogenous uncertainty (EXU & EDU) description is presented for four typical types of GES. Moreover, risk indices are proposed to assess the impact of overlooking (EXU & EDU) of GES. Comparative results between EXU & EDU are illustrated in distribution system with day-ahead chance-constrained optimization (CCO) and more severe risks are observed for the latter, which indicate that system operator (SO) should adopt novel strategies for EDU uncertainty.

## I. INTRODUCTION

HIGHER penetration of intermittent renewable energy resources (RES) raise increasing requirements to decline operational risk and promote system flexibility [1] from various flexible resources e.g., battery energy storage (BES), electric vehicle (EV), demand response (DR), etc. Part of these flexible resources have the attributes and abilities of ES, which are collectively called virtual energy storage (VES) [2]. The literature is rich especially in the modeling and optimization of VES. DR resources were divided into three types of VES, i.e., shiftable load, reducible load, and interruptible load according to their schedulable potential [3]. However, these DR models failed to maintain the consistency of BES model. A thermal battery storage model based on thermostatically controlled load was proposed in [4], which realized the conversion through 1st equivalent thermal parameter (ETP) model. And ref. [5] further extend it to district heating networks and buildings. However, the probabilistic properties are overlooked in these VES models. Different from real ES, VES unit usually has its own baseline usage with time-varying and stochastic processes (e.g., arrival and departure time of EV, temperature preference of TCL). Ref. [6] further investigated the uncertainties in model reduction error and uncontrollable attributes and formulate a CCO problem. The uncertainties are evaluated for the aggregation of DR resources and flexibility capacity-duration-probability curves are constructed to allow a VES operator to bid regarding uncertain capacity [7]. However, uncertainties considered in previous works are all exogenous, which is independent of operation strategies. While, EDU (e.g., response willingness, state of charge (SOC) preference) have been overlooked and can be altered with operation strategies and incentive prices.

Regarding risk assessment, reliability or risk indices, e.g., loss of load probability (LOLP), expected power not served (EENS), etc. are commonly used to evaluate the expectation of reliability or risk from the fault or outage of conventional electrical equipment and the volatility of RES. These indices are expected values extracted from determined probability distribution of these uncertain resources without VES. However, VES can't guarantee 100% reliable service, especially when overlooking the EDU of VES for power system risk assessment will lead to over-optimistic results and severe operational risks will occur unknown by SO though adopting risk-aware & risk-averse strategies, e.g., using CCO or robust optimization [6, 8].

Motivated by this background, we introduce the concept of our proposed GES and risk assessment for CCO under EXU and EDU. The main contributions of this paper are as follows.

1) GES is redefined, and a data-driven unified model is proposed for GES including four typical types, i.e., BES, inverter air-conditioner (IVA), fix frequency air-conditioner (FFA), and EV. More constraints and requirements are added to be more suitable for power system operation while guaranteeing both data privacy and parameter identifiability.

2) Uncertainty description of GES concerning response reliability and response capacity, EXU & EDU is presented, rendering a probabilistic model considered in CCO.

3) Risk indices are proposed to evaluate the predicted risk when overlooking the EXU & EDU of GES during CCO in distribution system. The case study illustrates that both EXU & EDU have negative impact on system operation and SO should adopt novel strategies considering EDU.

The remainder of this paper is organized as follows. The unified model and probability description are proposed in Section II. Both the approach of CCO and risk assessment under EXU & EDU are presented in Section III. Numerical analyses based on a real data-driven approach are provided in Section IV. Finally, the conclusions are summarized in Section V.

## II. PROBABILISTIC MODEL OF GENERIC ENERGY STORAGE

### A. Definition and classification of generic energy storage

Researchers in [9] define generic energy storage (GES) as any device with the capacity of transforming and storing energy. However, most of these flexible energy resources only have a one-way power transformation capacity. In addition, some restrictions are unnecessary to generate a general model, such as no stored energy losses, no hysteresis in charging and

N. Qi, L. Cheng, Y. Wan and Y. Zhuang are with State Key Laboratory of Control and Simulation of Power Systems and Generation Equipment, Department of Electrical Engineering, Tsinghua University, 100084 Beijing, China. e-mail: qn18@mails.tsinghua.edu.cn. Z. Liu is with Tianjin University.

This paper was sponsored by National Key R&D Program of China (Grant No. 2018YFE0116400), project of National Natural Science Foundation of China (Grant No. 52037006 & No. 51807107) discharging, etc. Therefore, to be more applicable, a GES device unit is redefined with the following assumptions.

1) Any GES unit should both have the capacity of charging or discharging. One-way transformation (only charging or discharging) is not qualified.

2) Stored energy losses are allowed. For instance, losses can happen due to self-discharge of BES or heat dissipation of TCL.

3) There will be hysteresis in charging or discharging, i.e., the time delay between response order and response action. But this dynamic process can be simplified into uncertainty when just considering the operation problem of GES.

4) Storage devices have conversion losses, i.e., charge or discharge efficiency. However, efficiency may be time-varying due to changing conditions.

5) Up and down ramps should be considered for the basic requirement of the power system. None of the units can go from idle state to full power state instantly.

6) Energy charged or discharged occurs at constant or mean value for a single time step, i.e., without consideration of short-time scale dynamics.

GES units can be divided into three types concerning the power and energy capacity, i.e., UPS for power quality enhancement, grid support and load shifting, bulk power migration. And they are divided as the ES and VES according to their probabilistic properties. VES units are used by aggregation and coordination because of their small capacity and uncertain properties, while ES units are more controllable.

*B. Data-driven unified model of generic energy storage*

To be specific, we focus on the economic dispatch of four types of the most commonly used GES, i.e., BES, IVA, FFA, and EV. For objective, we only consider incentive cost for GES shown in (1). $P_{\text{ch},i,t}^{\text{GES}}$ and $P_{\text{dis},i,t}^{\text{GES}}$ are decision variables of charge and discharge power, respectively. $c_{\text{ch},i,t}^{\text{GES}}$ and $c_{\text{dis},i,t}^{\text{GES}}$ are prices of charge and discharge power, respectively. Deterministic and chance constraints are generated as (2-8). Constraints (2-3) limit the upper and lower charging and discharging power. Constraint (4) limits the upper and lower SOC. Constraint (5) defines the relationship among charging power, discharging power, SOC, and output from ambient space. Constraint (6) limits the up and down ramping. Constraint (7) defines the power balance across time. The complementarity constraint $P_{\text{ch},i,t}^{\text{GES}} P_{\text{dis},i,t}^{\text{GES}} = 0$ can be relaxed and removed from the model.

$$C^{\text{GES}} = \sum_{t \in \Omega_T} \sum_{i \in \Omega_G} (P_{\text{dis},i,t}^{\text{GES}} c_{\text{dis},i,t}^{\text{GES}} + P_{\text{ch},i,t}^{\text{GES}} c_{\text{ch},i,t}^{\text{GES}}) \Delta t \quad (1)$$

**Constraints:** $\forall t \in \Omega_T$, $\forall i \in \Omega_G$

$$0 \leq P_{\text{ch},i,t}^{\text{GES}},\ 0 \leq P_{\text{dis},i,t}^{\text{GES}} \quad (2)$$

$$\mathbb{P}(P_{\text{ch},i,t}^{\text{GES}} \leq P_{\text{ch},i,\max}^{\text{GES}}) \leq 1-\alpha,\ \mathbb{P}(P_{\text{dis},i,t}^{\text{GES}} \leq P_{\text{dis},i,\max}^{\text{GES}}) \leq 1-\alpha \quad (3)$$

$$\mathbb{P}(SOC_{i,t,\min}^{\text{GES}} \leq SOC_{i,t}^{\text{GES}} \leq SOC_{i,t,\max}^{\text{GES}}) \leq 1-\alpha \quad (4)$$

$$SOC_{i,t+1}^{\text{GES}} = (1-\varepsilon_i^{\text{GES}}) SOC_{i,t}^{\text{GES}} + \eta_{\text{ch},i}^{\text{GES}} P_{\text{ch},i,t}^{\text{GES}} \Delta t / S_i^{\text{GES}}$$
$$- P_{\text{dis},i,t}^{\text{GES}} \Delta t / S_i^{\text{GES}} \eta_{\text{dis},i}^{\text{GES}} + \varepsilon_i^{\text{GES}} (SOC_{i,0}^{\text{GES}} - \beta_{i,t}^{\text{GES}}) \quad (5)$$

$$-Ramp_{i,\text{down}}^{\text{GES}} \Delta t \leq P_{i,t}^{\text{GES}} - P_{i,t-1}^{\text{GES}} \leq Ramp_{i,\text{up}}^{\text{GES}} \Delta t \quad (6)$$

$$SOC_{i,T}^{\text{GES}} = SOC_{i,0}^{\text{GES}} \quad (7)$$

Where $\Omega_T$ and $\Omega_G$ are sets of time periods and GES units, respectively. $SOC_{i,t}^{\text{GES}}$ is decision variable of SOC. $\Delta t$ is time step index, generally on an hourly basis. $T$ is the whole period of ED. For random variables, $P_{\text{ch},i,\max}^{\text{GES}}$ and $P_{\text{dis},i,\max}^{\text{GES}}$ are the maximum charge and discharge power, respectively. $SOC_{i,t,\max}^{\text{GES}}$ and $SOC_{i,t,\min}^{\text{GES}}$ are the maximum and minimum SOC boundaries, respectively. $\eta_{\text{ch},i}^{\text{GES}}$ and $\eta_{\text{dis},i}^{\text{GES}}$ are the charge and discharge efficiency, respectively. $\varepsilon_i^{\text{GES}}$, $S_i^{\text{GES}}$, $\beta_{i,t}^{\text{GES}}$ are the self-discharge rate and energy capacity, output from ambient space, respectively. $Ramp_{\text{up}}^{\text{GES}}$ and $Ramp_{\text{down}}^{\text{GES}}$ are the up and down ramping rate of GES.

Afterward, the relationship is analyzed between model's parameters and the physical parameters of each type summarized in Table 1. $C$, $R$, and $\eta$ are thermal capacity, thermal resistance, and conversion efficiency of TCL, respectively. The transformation of TCL into GES origins from thermal dynamics of 1st order ETP model. $T_{i,t}^{\text{in}}$, $T_{i,t}^{\text{out}}$ and $T_{i,t}^{\text{set}}$ are the indoor, outdoor, and setpoint temperature. These parameters can be obtained by some data-driven approach [10]. Great differences can be witnessed within these four types. For instance, the self-discharge rate is usually ignored for BES, but VESs tend to embrace a relatively high value. Ramping rate of BES and EV is restricted with maximum charge/discharge ratio, while it doesn't exist for TCLs, which is particularly helpful to improve the ramping capacity of power system. Besides, most of the parameters are relatively constant for BES and EV, but time-varying features appear for TCLs.

TABLE I COMPARISONS BETWEEN FOUR TYPES OF GES UNITS

| GES Type | BES | IVA | FFA | EV |
|---|---|---|---|---|
| $SOC_t^{\text{GES}}$ | $SOC_t^{\text{BES}}$ | $\dfrac{T_{\max}^{\text{in,IVA}} - T_t^{\text{in,IVA}}}{T_{\max}^{\text{in,IVA}} - T_{\min}^{\text{in,IVA}}}$ | $\dfrac{T_{\max}^{\text{in,FFA}} - T_t^{\text{in,FFA}}}{T_{\max}^{\text{in,FFA}} - T_{\min}^{\text{in,FFA}}}$ | $SOC_t^{\text{EV}}$ |
| $P_{\text{ch,max}}^{\text{GES}}$ | $P_{\text{ch,max}}^{\text{BES}}$ | $\dfrac{P_{\max}^{\text{IVA}} - \dfrac{T_0^{\text{out,IVA}} - T_0^{\text{set,IVA}}}{\eta^{\text{IVA}} R^{\text{IVA}}}}{}$ | $\dfrac{P_{\text{rated}}^{\text{FFA}} - \dfrac{T_0^{\text{out,FFA}} - T_0^{\text{set,FFA}}}{\eta^{\text{FFA}} R^{\text{FFA}}}}{}$ | $P_{\text{ch,max}}^{\text{EV}}$ |
| $P_{\text{dis,max}}^{\text{GES}}$ | $P_{\text{dis,max}}^{\text{BES}}$ | $\dfrac{T_0^{\text{out,IVA}} - T_0^{\text{set,IVA}}}{\eta^{\text{IVA}} R^{\text{IVA}}}$ | $\dfrac{T_0^{\text{out,FFA}} - T_0^{\text{set,FFA}}}{\eta^{\text{FFA}} R^{\text{FFA}}}$ | $P_{\text{dis,max}}^{\text{EV}}$ |
| $SOC_{t,\min}^{\text{GES}}$ | $SOC_{\min}^{\text{BES}}$ | $\dfrac{T_{\max}^{\text{in,IVA}} - T_{t,\max}^{\text{in,IVA}}}{T_{\max}^{\text{in,IVA}} - T_{\min}^{\text{in,IVA}}}$ | $\dfrac{T_{\max}^{\text{in,FFA}} - T_{t,\max}^{\text{in,FFA}}}{T_{\max}^{\text{in,FFA}} - T_{\min}^{\text{in,FFA}}}$ | $SOC_{\min}^{\text{EV}}$ |
| $SOC_{t,\max}^{\text{GES}}$ | $SOC_{\min}^{\text{BES}}$ | $\dfrac{T_{\max}^{\text{in,IVA}} - T_{t,\min}^{\text{in,IVA}}}{T_{\max}^{\text{in,IVA}} - T_{\min}^{\text{in,IVA}}}$ | $\dfrac{T_{\max}^{\text{in,FFA}} - T_{t,\min}^{\text{in,FFA}}}{T_{\max}^{\text{in,FFA}} - T_{\min}^{\text{in,FFA}}}$ | $SOC_{\max}^{\text{EV}}$ |
| $\varepsilon^{\text{GES}}$ | $\varepsilon^{\text{BES}}$ | $1 - e^{-\Delta t / R^{\text{IVA}} C^{\text{IVA}}}$ | $1 - e^{-\Delta t / R^{\text{FFA}} C^{\text{FFA}}}$ | $\varepsilon^{\text{EV}}$ |
| $S^{\text{GES}}$ | $S^{\text{BES}}$ | $\dfrac{\Delta t(T_{\max}^{\text{in,IVA}} - T_{\min}^{\text{in,IVA}})}{\eta^{\text{IVA}} R^{\text{IVA}} \varepsilon^{\text{IVA}}}$ | $\dfrac{\Delta t(T_{\max}^{\text{in,IVA}} - T_{\min}^{\text{in,IVA}})}{\eta^{\text{FFA}} R^{\text{FFA}} \varepsilon^{\text{FFA}}}$ | $S^{\text{EV}}$ |
| $\beta^{\text{GES}}$ | $SOC_0^{\text{BES}}$ | $\dfrac{T_{t+1}^{\text{out,IVA}} - T_0^{\text{out,IVA}}}{T_{\max}^{\text{in,IVA}} - T_{\min}^{\text{in,IVA}}}$ | $\dfrac{T_{t+1}^{\text{out,FFA}} - T_0^{\text{out,FFA}}}{T_{\max}^{\text{in,IVA}} - T_{\min}^{\text{in,IVA}}}$ | $SOC_0^{\text{EV}}$ |
| $\eta_{\text{ch}}^{\text{GES}}$ | $\eta_{\text{ch}}^{\text{BES}}$ | 1 | 1 | $\eta_{\text{ch}}^{\text{EV}}$ |
| $\eta_{\text{dis}}^{\text{GES}}$ | $\eta_{\text{dis}}^{\text{BES}}$ | 1 | 1 | $\eta_{\text{dis}}^{\text{EV}}$ |
| $Ramp_{\text{up}}^{\text{GES}}$ $Ramp_{\text{down}}^{\text{GES}}$ | $Ramp_{\max}^{\text{BES}}$ | — | — | $Ramp_{\max}^{\text{EV}}$ |

### C. Uncertainty description of generic energy storage

Notably, GES is not entirely controllable by SO for flexible DR projects. After issuing DR orders, consumers can choose both the response probability and capacity for each period, which will be a trade-off strategy between their expected earning and inconvenience cost and we further discuss it in [11].

**(a) Response probability:** it's the discrete uncertainty with binary status to evaluate the working or operating state of GES. As for ES, it means reliability and security, while it refers to operating rate and response willingness for VES.

**(b) Response capacity:** it's the continuous uncertainty to evaluate real-time response power of GES constricted with four main aspects. (b1) SOC boundary preference: SOC boundaries are with EXU, e.g., temperature preference of TCL, SOC preference for multi-services of ES and EV. Also, SOC boundaries will perform expansion and contract effect due to incentive price and accumulated discomfort, rending EDU. (b2) Parameter identification errors: due to data sampling error and simplification of high-resolution dynamics. (b3) Forecast errors: temperature forecast error will add uncertainty input to Eq. (5). (b4) Capacity degradation: it's a nonlinear and complex process for ES, but it means the decline of response units for VES.

## III. CHANCE-CONSTRAINED OPTIMIZATION AND RISK ASSESSMENT UNDER EXU & EDU

### A. Problem Formulation with CCO

Hereby, we consider the voltage management in distribution system shown in Fig.1. Previously, SO will adopt optimal load curtailment strategies to handle the low-voltage problem in the distribution system considering the uncertainties from grid, load, RES, and conventional power plant (CPP), where ES is treated as 100% reliable devices to improve system's economy and reliability. However, if considering the EXU & EDU of GES, the problem is further formulated as follows with CCO.

$$\min f(\boldsymbol{x},\boldsymbol{\xi}) = \sum_{t\in\boldsymbol{\Omega}_T}\sum_{i\in\boldsymbol{E}_n} P_{i,t}^{\text{LC}}\Delta t + \sum_{t\in\boldsymbol{\Omega}_T} c_t^{\text{Grid}} P_t^{\text{Grid}} + C^{\text{GES}} \quad (8)$$

**Constraints:** $\forall t \in \boldsymbol{\Omega}_T$, $i \in \boldsymbol{E}_n$, $ij \in \boldsymbol{E}_b$

$$\mathbb{P}(\sum_{ik\in\boldsymbol{E}_b} P_{ik,t} + P_{i,t}^{\text{RES}} + \sum_{i\in\boldsymbol{\Omega}_G}(P_{\text{dis},i,t}^{\text{GES}} - P_{\text{ch},i,t}^{\text{GES}}) \\ - \sum_{ji\in\boldsymbol{E}_b}\left(P_{ji,t} - r_{ij}I_{ji,t}\right) \geq P_{i,t}^{\text{L}} - P_{i,t}^{\text{LC}}) \geq 1-\alpha \quad (9)$$

$$\mathbb{P}(\sum_{ik\in\boldsymbol{E}_b} Q_{ik,t} + Q_{i,t}^{\text{RES}} + \sum_{s\in\boldsymbol{\Omega}_G}(Q_{\text{dis},s,t}^{\text{GES}} - Q_{\text{ch},s,t}^{\text{GES}}) \\ - \sum_{ji\in\boldsymbol{E}_b}\left(Q_{ji,t} - x_{ij}I_{ji,t}\right) \geq Q_{i,t}^{L} - Q_{i,t}^{\text{LC}}) \geq 1-\alpha \quad (10)$$

$$U_{i,t} - U_{j,t} + \left(r_{ij}^2 + x_{ij}^2\right)I_{ij,t} - 2\left(r_{ij}P_{ij,t} + x_{ij}Q_{ij,t}\right) = 0 \quad (11)$$

$$U_i^{\min} \leq U_{i,t} \leq U_i^{\max}, \quad 0 \leq I_{ij,t} \leq I_{ij}^{\max} \quad (12)$$

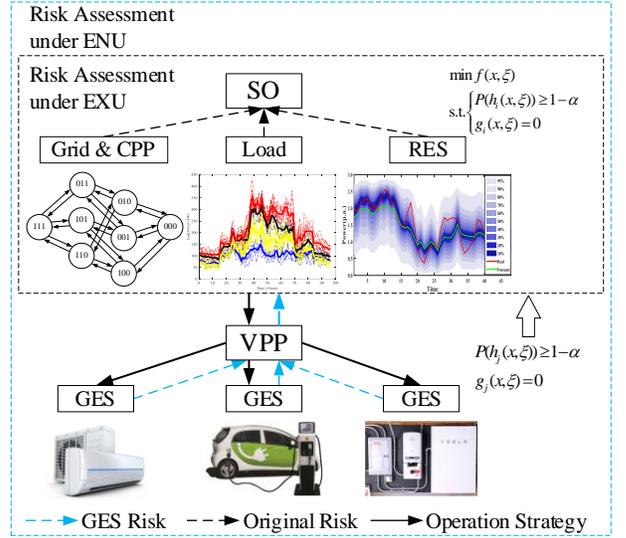

Fig. 1. Comparison of risk assessment under EXU & EDU

$$\left\|\begin{bmatrix} 2P_{ij,t} & 2Q_{ij,t} & I_{ij,t} - U_{i,t} \end{bmatrix}^T\right\|_2 \leq I_{ij,t} + U_{i,t} \quad (13)$$

where (8) is the overall cost considering load curtailment, incentive cost of GES, electricity cost from the grid. Constraints (9-10) are chance constraints of power balance. Constraints (11-13) are power flow constraints using disflow model and second-order cone relaxation. $P_t^{\text{Grid}}$, $P_{i,t}^{\text{LC}}$, $P_{i,t}^{\text{RES}}$ and $P_{i,t}^{L}$ are the grid power, active power of load curtailment (PLC), RES power and load power, respectively, and it's the same with Q for reactive power. $P_{ij,t}$, $I_{ij,t}$, $r_{ij}$ and $x_{ij}$ are active power, current, resistance and reactance of branch. $\boldsymbol{E}_n$ and $\boldsymbol{E}_b$ are the sets of nodes and branches. $1-\alpha$ is the security level. $c_t^{\text{Grid}}$ is the electricity price bought from the upper grid.

### B. Risk assessment under EXU & EDU

Notably, system's risks are within security level using CCO, however, things will be different if the EXU & EDU of GES are overlooked by SO. The reason for overlooking is that first, it's hard for SO to obtain detailed EXU analysis of GES due to data privacy. On another hand, the EDU (also called decision-dependent uncertainty) will be affected by strategies and can not be fully determined before operation. Thus, we introduce two reliability indices, i.e., loss of response probability (LORP) and expected response power not severed (ERNS) to assess the reliability losses for overlooking the uncertainty of GES in CCO and they are described as follows, $X_k | \boldsymbol{x},\boldsymbol{\xi}$ represents the reliability loss events under strategy $\boldsymbol{x}$ and uncertainty $\boldsymbol{\xi}$. $E(\cdot)$ is the function of response energy losses.

$$\begin{cases} LORP = \sum_k \mathbb{P}(X_k | \boldsymbol{x},\boldsymbol{\xi}) \\ ERNS = \sum_k \mathbb{P}(X_k | \boldsymbol{x},\boldsymbol{\xi})E(X_k | \boldsymbol{x},\boldsymbol{\xi}) \end{cases} \quad (14)$$

## IV. NUMERICAL ANALYSIS

### A. Set-up

CCO and risk assessment are analyzed in IEEE 33 bus bench-

-mark test system. Historical data of RES and load are collected from low voltage urban area of Jiangsu province, China in 2020, where normalized value is used as input. Wind and solar generators are located on bus 7, 24, 25 and 32 which are with the highest load level. Each RES is with 300 kW rated capacity. GES system with 100 TCL units is located on bus 1 to support grid operation. Parameters and data can be obtained from real households in the Muller project in Austin, TX, USA. All the data used for this combined system can refer to [12].

### B. Optimization performance with different uncertainties

We compute and compare the results for dispatch strategies and voltage performance between three models with different uncertainties.

**(M1) Without considering uncertainties of GESs.** GESs are set with deterministic parameters (i.e., mean value), load and RES are set with random parameters.

**(M2) Considering EXU of GESs.** Based on **(M1)**, GESs parameters are changed with random ones.

**(M3) Considering EXU & EDU of GESs.** Based on **(M2)**, EDU of GESs shown in [11] are added into constraints.

For **(M2)** and **(M3)**, $\alpha$ is set to be 0.05. The price for load curtailment is set to be 10 RMB/kW, incentive price for GESs is 0.3 RMB/kW. We found that the voltage performance and PLC are maintained the same for different models, the results shown in Fig. 2 illustrates that voltage is controlled within the security domain with certain PLC. And GESs may not reduce the PLC because the essence of load loss is insufficient generation adequacy. Although using GESs can realize energy transfer across time, it will further aggravate the average energy gap due to the energy losses caused by GESs themselves. However, they are helpful to reduce the cost because they are less costly compared with electricity cost and PLC cost. By comparison shown in Table I, as the uncertainties considered increase, the optimum cost is gradually getting worse and the usage of GES declines. Moreover, the comparison of GES strategies shown in Fig. 3 also indicates that the average SOC level (bold lines) is increasing especially considering EXU, and GESs tend to discharge less at the beginning and charge more at the second half. The results are more conservative because the feasible region is reduced when considering EXU&EDU.

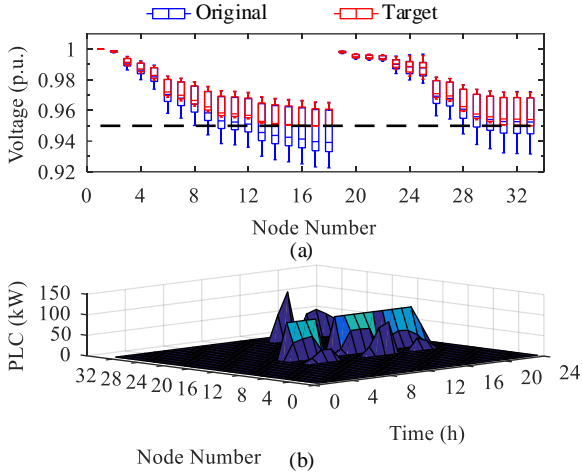

Fig. 2. Comparison of (a) voltage distribution and (b) PLC distribution

TABLE I OPTIMIZATION RESULTS COMPARED WITH DIFFERENT MODELS

| Output | M1 | M2 | M3 |
| --- | --- | --- | --- |
| Cost (RMB) | 66714.96 | 66849.00 | 66948.65 |
| $\sum P_{dis,i,t}^{GES}$ (kW) | 413.84 | 330.24 | 79.62 |
| $\sum P_{ch,i,t}^{GES}$ (kW) | 306.60 | 369.13 | 354.92 |
| $\sum P_t^{Grid}$ (kW) | 41827.96 | 41974.12 | 42210.58 |
| $\sum P_{i,t}^{LC}$ (kW) | 2359.05 | 2359.05 | 2359.05 |
| $\overline{U}_{i,t}$ (p.u.) | 0.97 | 0.97 | 0.97 |

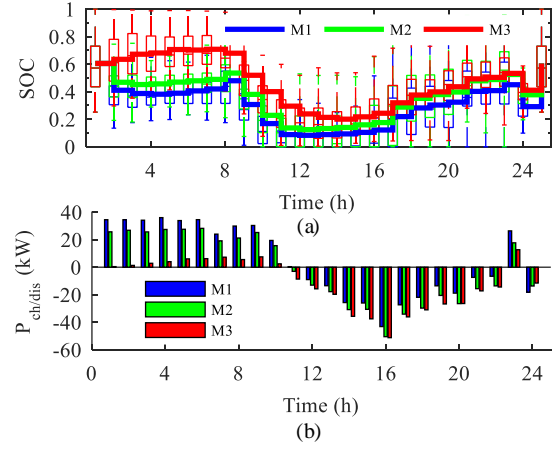

Fig. 3. Comparison of GES strategies (a) SOC and (b) charge/discharge power

### C. Comparison between EXU & EDU

In this subsection, the SOC boundaries are compared between EXU & EDU to explain the different optimization results discussed above. SOC boundaries represent time-varying setpoint temperature for TCL-GESs. Previously, the SOC boundaries are with EXO described as blue lines shown in Fig. 4. The boxplot is the distribution of 100 TCL-GESs' SOC boundaries, boundaries with security between 0.05-0.95 are shown within cyan lines, and the dark blue lines with the security of 0.5 are used for deterministic parameters in M1. The feasible region declines when the security level raised from 0.5 (M1) to 0.95 (M2). Thus, the results of M2 are more conservati-

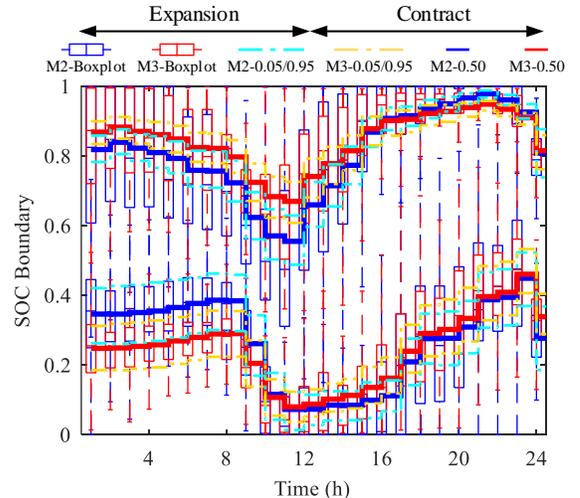

Fig. 4. Comparison of SOC boundaries under EXU & EDU

-ve than M1. Additionally, expansion effect appears described as red and orange lines when considering EDU. It's observed that the feasible region is enlarged and reduced during the two periods. What interests us most is that the decisions of GESs choose to respond less in the beginning and reserve more feasible region for the second half because the boundaries will be reduced due to the accumulated discomfort. Thus, strategies of GESs are more conservative to consider EDU, and it can also well explain the refused response and load recovery in DR.

*D. Risk assessment with different uncertainties*

Although results in M1&M2 outperform M3 with respect to optimality, they seem not quite reliable, that is to say, GESs may not act as the same with strategies. Thus, the purpose of our paper is to assess a novel type of risk (i.e., risk of response), rather than the general risks caused by RES and load. Fig. 5 (a) illustrates the difference between practical (rainbow color) and theoretical (red color) SOC boundaries with different security levels in M2 (γ=0.5 for M1). It is observed that the practical upper and lower boundaries will gradually compress into the same value (central value). And the difference between the practical and theoretical ones is the main reason for losses of response reliability. Reliability performance shown in Table II indicates the practical security level will exceed the setpoint thus causing unpredicted voltage problems and extra PLC. Moreover, the risk of M2 increase dramatically using strategies with the decline of security level. However, the results of M3 outperform its rivals in both LORP and ERNS indices, which guarantee the risks within the security level.

## V. CONCLUSION

In this paper, we introduce the concept and data-driven unified model of GES. Detailed uncertainty descriptions about EXU&EDU of GES are presented, additionally, they are fully considered in both CCO problem and risk assessment. Conclusions from this work are listed below:

1) Considering EDU of incentive price and accumulated discomfort on SOC boundaries of GES enables decision-makers to consider the negative impact of strategies on feasible regions and produce more conservative strategies by capturing a trade-off between profitability and net comfort of customers.

2) In terms of reliability performance, the proposed optimization model outperforms the previous models and offers nearly 100% reliable strategies which reduce the requirement for real-time power balance and costs for the reserve. The main reasons are that 1) reduced incomplete knowledge via data-driven work. 2) consideration of EDUs on the feasible region.

3) Rather than the general risks from RES and load, the risk assessment proposed in this paper focus on the risk of real response and actions of GES, which provides reliable consequence analysis about unpredictable risks caused by EDUs and prevent the over-optimistic strategies made by decision-maker when using the probabilistic devices like GES.

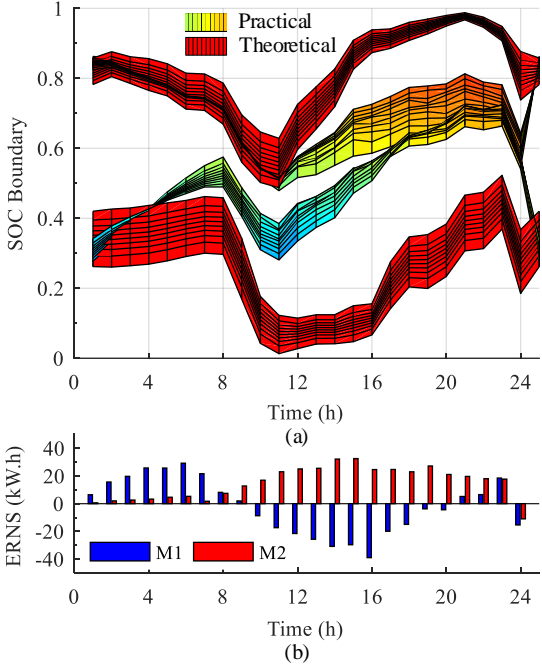

Fig. 5 Reliability performance comparison of (a) SOC boundaries (b) ERNS

TABLE II RISK ASSESSMENT COMPARED WITH DIFFERENT MODELS

| γ | Index | M1 | M2 | M3 |
|---|---|---|---|---|
| 0.05 | LORP | LORP 0.907 | 0.745 | 0.022 |
|  | ERNS |  | 15.940 | 0.036 |
| 0.25 | LORP |  | 0.813 | 0.249 |
|  | ERNS | ERNS 25.327 | 20.208 | 0.703 |
| 0.45 | LORP |  | 0.863 | 0.439 |
|  | ERNS |  | 23.161 | 0.840 |

REFERENCES

[1] F. L. Mueller, S. Woerner and J. Lygeros, "Unlocking the Potential of Flexible Energy Resources to Help Balance the Power Grid," *IEEE Transactions on Smart Grid*, vol. 10, no. 5, pp. 5212-5222, Sep. 2019.
[2] A. Niromandfam, A. M. Pour, and E. Zarezadeh, "Virtual energy storage modeling based on electricity customers' behavior to maximize wind profit," *Journal of Energy Storage*, vol. 32, pp. 101811, Dec. 2020.
[3] Y. Xia, Q. Xu, H. Qian, et al, "Bilevel optimal configuration of generalized energy storage considering power consumption right transaction," *International Journal of Electrical Power & Energy Systems*, 2021, 128: 106750.
[4] M. Song, C. Gao, H. Yan, et al., "Thermal Battery Modeling of Inverter Air Conditioning for Demand Response," *IEEE Trans. on Smart Grid*, vol. 9, no. 6, pp. 5522-5534, Nov. 2018.
[5] S. Lu, W. Gu, K. Meng, et al, "Thermal Inertial Aggregation Model for Integrated Energy Systems," *IEEE Trans. on Power Systems*, vol. 35, no. 3, pp. 2374-2387, May 2020.
[6] M. Amini and M. Almassalkhi, "Optimal Corrective Dispatch of Uncertain Virtual Energy Storage Systems," *IEEE Trans. on Smart Grid*, vol. 11, no. 5, pp. 4155-4166, Sep. 2020.
[7] J. Zhang and A. D. Domínguez-García, "Evaluation of Demand Response Resource Aggregation System Capacity Under Uncertainty," *IEEE Trans. on Smart Grid*, vol. 9, no. 5, pp. 4577-4586, Sep. 2018.
[8] C. Ordoudis, V. A. Nguyen, D. Kuhn, et al, "Energy and reserve dispatch with distributionally robust joint chance constraints," *Operations Research Letters*, vol. 49, no. 3, pp. 291–299, May 2021.
[9] D. Pozo, J. Contreras and E. E. Sauma, "Unit Commitment With Ideal and Generic Energy Storage Units," *IEEE Trans. on Power Systems*, vol. 29, no. 6, pp. 2974-2984, Nov. 2014.
[10] N. Qi, L. Cheng, H. Xu, et al, "Smart meter data-driven evaluation of operational demand response potential of residential air conditioning loads," *Applied Energy*, vol. 279, no. 1, pp. 115708, Dec. 2020.
[11] N. Qi, P. Pinson, L. Cheng, et al, "Chance Constrained Economic Dispatch of Generic Energy Storage under Decision-Dependent Uncertainty," 2022. [Online]. Available: https://arxiv.org/abs/2201.06407.
[12] Supporting data brief [Online]. Available: https://data.mendeley.com/data sets/5vvffh53r5/2.